# Support of Teleoperated Driving with 5G Networks

M.C. Lucas-Estañ[1], B. Coll-Perales[1], M. I. Khan[2], S. S. Avedisov[2], O. Altintas[2], J. Gozalvez[1], M. Sepulcre[1]
[1]Uwicore laboratory, Universidad Miguel Hernandez de Elche, Elche (Alicante), Spain.
[2]InfoTech Labs, Toyota Motor North America R&D, Mountain View, CA, U.S.A.
{m.lucas, bcoll, j.gozalvez, msepulcre}@umh.es; {mohammad.irfan.khan, sergei.avedisov, onur.altintas}@toyota.com

***Abstract*—Teleoperated driving (ToD) can support autonomous driving under complex or unexpected traffic scenarios that an autonomous vehicle may not understand or be able to handle. In ToD, autonomous vehicles transmit video feeds and perception data to the remote control center. The operator uses this data to understand the driving environment and remotely control the vehicle that can take over the control once the scenario is resolved. ToD requires reliable and low latency communications between the vehicle and the ToD control center. This study analyzes the feasibility to support ToD with 5G networks. The study demonstrates that the feasibility strongly depends on the bandwidth and the Time Division Duplexing (TDD) frame structure that conditions how the bandwidth is distributed between uplink and downlink transmissions. The study also shows that scaling the number of 5G-supported ToD vehicles requires the vehicles to reduce the video bitrates. The study also shows that traditional centralized 5G network deployments may be challenged by some of the most stringent ToD latency requirements due to the latency introduced by the Internet connection to the ToD control center.**

## I. Introduction

Autonomous vehicles may encounter challenging or unexpected traffic conditions outside their operational design domains that they cannot understand or resolve safely or efficiently. Teleoperated driving (ToD) can help overcome these challenges and support autonomous driving in critical or delicate scenarios. With ToD, an operator can take over the control of vehicles from a ToD control center and drive them remotely. The operator can drive a vehicle in real-time or just provide a driving path for the vehicle to follow in order to resolve the challenging scenario. ToD requires autonomous vehicles to transmit video feeds and perception data collected from its onboard cameras and sensors to the ToD control center in order to provide the operator with real-time and accurate information about the vehicle's surroundings. The operator uses this information to understand the driving environment and safely control the vehicle from the remote ToD control center. This operation requires low latency, reliable and potentially bandwidth-demanding (in particular for the uplink) wireless connections between the vehicle and the remote ToD control center for a safe and effective ToD operation.

5G networks can provide the low latency, high reliability and bandwidth demands of ToD. The capacity of 5G to support ToD has been recently explored in European research projects such as 5GCroCo (https://5gcroco.eu/) and 5GBluePrint (https://www.5gblueprint.eu/). These projects focus on field trials in pilot deployments involving a limited number of ToD vehicles (often just one vehicle) in common 5G operational scenarios. For example, trials in 5GcroCo involved a centralized 5G network with 40 MHz in the n78 band at 3.7 GHz and a 4:1 Time Division Duplexing (TDD) frame structure consisting of 4 slots for the downlink (DL) and 1 slot for the uplink (UL). The ToD control center was accessible through the Internet, and trials involved a single vehicle transmitting video at different bitrates. The study in [1] considers a 2x20 MHz Frequency Division Duplexing (FDD) 5G network with the ToD control center using a dedicated 10 Gb/s wired connection to the mobile network. The simulation-based study focuses again on a single ToD vehicle transmitting video at 32 Mbps (following 5G Automotive Associate -5GAA- ToD requirements in [2]) but considers additional background traffic generated by other vehicles. The capabilities of 4G Long Term Evolution (LTE) to support ToD have also been studied in [3]. The paper shows that the limitations of LTE in supporting ToD regarding latency can be mitigated by adapting the video frame rate.

Existing studies provide important insights into the capability of 5G to support ToD. However, it is still necessary to investigate whether the ToD service can scale in 5G networks. The 5GAA identifies a target density of ToD vehicles of 10 veh/km$^2$ [2], but it is unclear if 5G networks can support it, and the configurations under which it could. In fact, current 5G networks mostly rely on centralized deployments with TDD frame structures that allocate more radio resources for the DL than the UL since current mobile services are still highly asymmetric (e.g. video streaming services). On the other hand, the ToD service puts more stress on the UL with the transmission of perception data (e.g., video feeds) while the DL is basically used to send driving commands from the ToD control center. In this context, this study advances the state-of-the-art by analyzing the configurations and conditions under which 5G networks can support ToD at different vehicle densities. In particular, the main contributions of this paper are:

- We quantify the impact of different TDD frame structures and bandwidths on the capability of 5G to support the ToD service.
- We also show that adapting the resolution of video feeds transmitted by the vehicles to the ToD control center can improve the scalability of the ToD service.
- We highlight the challenge to support some of the stringent ToD service requirements with traditional centralized 5G network deployments due to the latency introduced by the Internet connection to the ToD control center.

## II. Tele-Operated Driving over 5G

The 5GAA is one of the most active initiatives on the study of ToD since it brings together the two main players for its operation, i.e., automotive and telecommunication industries. The 5GAA opened a Work Item on ToD aimed at identifying use cases and describing the requirements needed for ToD service provisioning. The 5GAA identifies four different ToD use cases [4]: 1) ToD; 2) ToD support; 3) ToD for automated parking; and 4) Infrastructure-based ToD. The ToD use case considers that an operator remotely drives a vehicle using perception data collected by the vehicle in order to understand the driving environment. The ToD support and ToD for automated parking use cases are restricted to the remote operation of the vehicle for a short period of time and in a confined area, respectively. The infrastructure-based ToD use case considers that the remote driver relies on perception data provided by sensors on the road infrastructure. All use cases consider that ToD is initiated after the vehicle has suffered an incident or finds itself in challenging scenarios (e.g., a maneuver it cannot complete autonomously) [5]. The remote operator can directly or indirectly control the vehicle. In the case of an indirect control, the remote operator does not directly drive the vehicle but provides guidelines or a path for the vehicle to overcome the encountered challenge; the vehicle controls the execution of the path. On the other hand, the remote operator fully controls the vehicle in the case of direct control. Direct control entails more stringent requirements on the 5G connectivity [5]. This is particularly the case for the ToD use case since a real-time actuation over the vehicle requires timely and reliable transmission of perception data and driving commands. This study focuses on the ToD use case with direct control, whose requirements are summarized in Table I.

5GAA considers that ToD vehicles should mount 4 cameras, and each camera generates 8 Mbps. This results in a total UL video bitrate per vehicle of 32 Mbps. 5GAA establishes a 100 ms service latency requirement and a 99% reliability requirement for UL transmissions [2] [1]. The reliability is defined as the percentage of packets successfully delivered within the time constraint required by the target service (100 ms for UL ToD transmissions) [6]. 5GAA imposes even more stringent requirements for the DL transmission of 1KB-command messages from the remote driver to the vehicle since they directly "*affect the safe and efficient operation of the vehicle*". 5GAA establishes that these commands should be sent every 20 ms (400 Kbps) with a 99.999% reliability requirement [2]. 5GAA also sets a target density of ToD vehicles equal to 10 veh/km$^2$.

TABLE I. ToD Service Level Requirements [2]

| **Data Rate** | **Service Level Latency** | **Reliability** | **Density** |
|---|---|---|---|
| UL: 32 Mbps<br>DL: 400 Kbps | UL: 100 ms<br>DL: 20 ms | UL: 99%<br>DL: 99.999% | 10 veh/km$^2$ |

## III. 5G Latency Modelling

This study analyses the capability of 5G to support the service latency and reliability requirements defined by 5GAA for ToD. Since UL and DL transmissions have different requirements and traffic demands, we study separately the latency experienced in the transmission of the data from the vehicle to the ToD control center (UL latency or $l_{UL}$), and from the ToD control center to the vehicle (DL latency or $l_{DL}$). We quantify the UL and DL latency using the models presented in [6] and [8]. We consider a traditional centralized 5G network deployment such as the one depicted in Fig. 1. In this case, the latency model accounts for the latency experienced at the radio, transport (TN) and core (CN) networks ($l_{radio}$, $l_{TN}$ and $l_{CN}$, respectively), and the latency introduced by the Internet connection between the last UPF (User Plane Function) in the CN and the V2X application server (AS) implemented in the ToD control center that is hosted on the cloud ($l_{UPF\text{-}AS}$). $l_{UL}$ and $l_{DL}$ are expressed as:

$$l_x = l_{radio} + l_{TN} + l_{CN} + l_{UPF\text{-}AS}, \text{ with } x=\{\text{UL, DL}\} \quad (1)$$

$l_{radio}$ is estimated using the model in [6]. This model considers the use of different 5G NR numerologies and Sub-Carrier Spacing (SCS), Modulation and Coding Schemes (MCSs), the use of full-slots or mini-slots, semi-static and dynamic scheduling, different retransmission mechanisms, and broadcast/multicast or unicast transmissions. The model was originally derived for FDD 5G networks and the transmission of short radio packets. The model is extended in this study to also model TDD 5G networks as well as the transmission of large packets generated by video cameras from ToD vehicles. Large packets may need to be segmented before the UL transmission over the radio channel. The TDD frame structure defines how slots are distributed between UL and DL. It then impacts the time a packet waits for radio resources before it can be transmitted on the UL or DL. The TDD frame structure also affects the latency introduced in the signalling processes for scheduling and Hybrid Automatic Repeat Request (HARQ) retransmissions.

$l_{TN}$ and $l_{CN}$ in (1) account for the propagation and transit delays over the TN and CN. The propagation delay depends on the length of the links and represents the time packets need to travel through the links at the TN or CN. The transit delay accounts for the time needed to receive, process, and transmit packets at TN or CN nodes. We use queueing theory to compute the transit delay that depends on the number of nodes packets pass through, the network load, and the capacity of the links. $l_{UPF\text{-}AS}$ in (1) is modelled using the empirical study in [9] that characterizes the round-trip time observed between source-target Internet nodes in the same country.

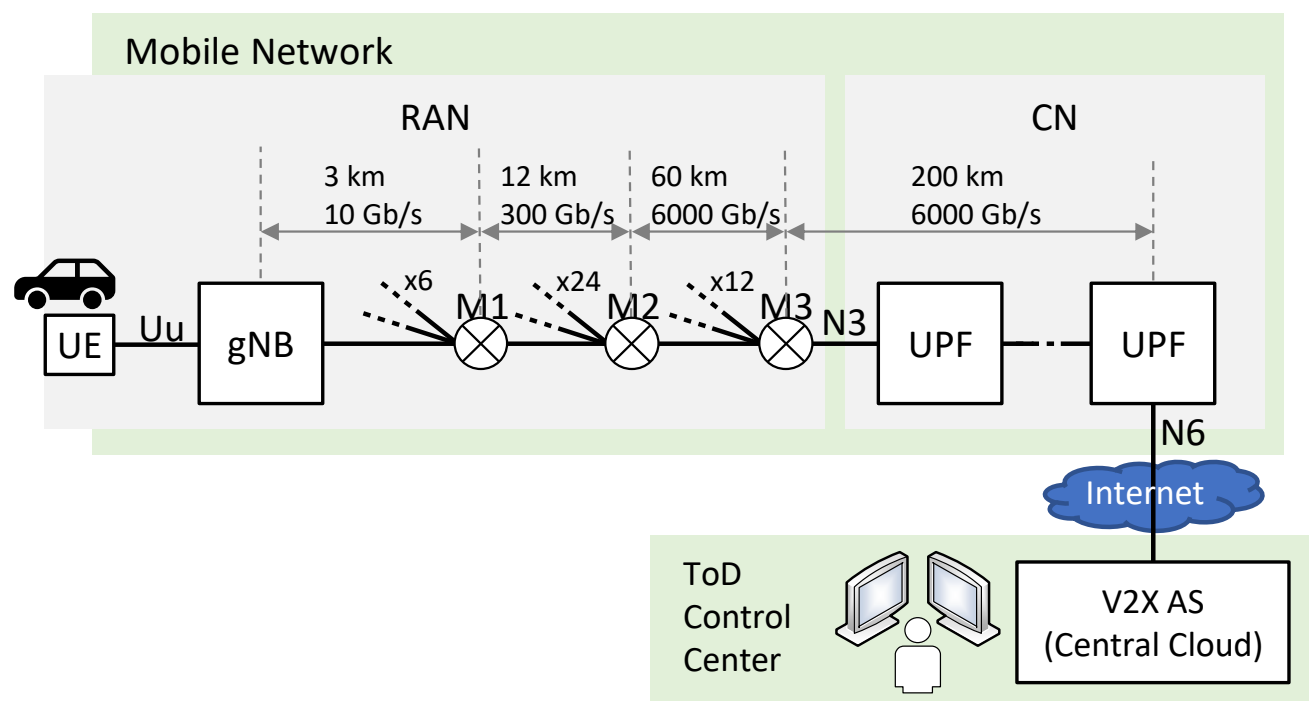

**Fig. 1**. 5G network with centralized deployment.

[1] In this study, we do not consider processing delays or human reaction times when analyzing the ToD service latency levels based on the stringent 5GAA requirements.

## IV. EVALUATION SCENARIO

We consider a single-cell 5G network that covers a 3-lane highway scenario. We analyze the feasibility to support the ToD service considering a single ToD vehicle per 5G cell, or several ToD vehicles per cell. 5GAA establishes that 5G networks should support a density of ToD vehicles up to 10 veh/km$^2$ [2]. This corresponds to a density between 1 and 2 veh/km/lane in our scenario [2], assuming all the vehicles requiring teleoperation are located on the highway. The baseline ToD configuration is that defined by 5GAA in [2] (see Section II). We also evaluate the impact of transmitting lower quality video based on the results of the ToD trials reported by European projects such as [10].

We consider a centralized 5G network deployment (Fig. 1) that follows the topology recommended by ITU in [11]. The network implements a hierarchical TN with 3 multiplexing nodes (M1, M2 and M3 in Fig. 1) that multiplex the traffic from 6 gNBs, 24 M1 and 12 M2 nodes, respectively. The distance and link capacities between any two nodes of the network are indicated in Fig. 1. We estimate and reserve for V2X traffic the fraction of the link capacities needed to avoid the backlog of V2X packets at TN or CN nodes. The V2X AS hosted at the ToD control center is implemented in the cloud and is accessible through an Internet connection.

We consider a 5G NR cell with a radius of 866 m [12]. 5G NR is configured with a SCS of 30 kHz, 2 Multiple-Input Multiple-Output (MIMO) layers, and full-slot transmissions. We consider semi-static scheduling for both UL and DL transmissions. Considering the 5GAA requirements for the ToD service, UL and DL transmissions use the MCSs defined in MCS Tables 2 and 3 in [13], and the MCS is adapted as a function of the Channel Quality Indicator (CQI) to achieve a target Block Error Rate (BLER) of 0.1 or $10^{-5}$, respectively. The UL is configured with a maximum of 3 HARQ retransmissions. We evaluate the capability of 5G to support the ToD service with two TDD frame structures that are recommended by GSMA in [14] and that are the most commonly used in current 5G deployments: DDDSU (labeled TDD1) and DDDDDDDSUU (TDD2), where D and U represent slots reserved for DL and UL transmissions respectively. S represents a special slot with a ratio of 10:2:2 (DDDSU) and 6:4:4 (DDDDDDDSUU) symbols reserved for DL, Guard Period and UL, respectively. Without loss of generality, our study assumes the S slot as a D slot given the larger number of DL symbols in S. We consider a third TDD frame structure (TDD3) with a more balanced distribution of slots for UL and DL: DUDU. This TDD frame structure is compatible with the 5G standard [15], but it is not widely deployed since current mobile services are still highly asymmetric (e.g. video streaming services).

We assume a baseline bandwidth (BW) of 40 MHz as it is frequently used in ToD trials [10] [16]. However, we also analyze the impact of the BW (30-100 MHz) on the capacity of 5G to support the ToD service. We also analyze the performance achieved with an FDD configuration given its latency benefits; in this case, BW/2 is reserved for UL and DL transmissions.

[2] $x$ veh/km/lane is equal to $x{\cdot}6$ veh/km$^2$ in our evaluation scenario because we consider a highway with 6 lanes and the total width of the highway is less than 1 km.

## V. IMPACT OF THE 5G CONFIGURATION ON THE TOD SERVICE

This section analyzes the impact of the 5G network configuration on the capacity to sustain the ToD requirements when considering a single ToD vehicle per cell. We separate the UL and DL evaluations since their bandwidth and ToD requirements significantly differ, and most current 5G network deployments asymmetrically allocate resources for UL and DL [10].

### *A. UL ToD traffic*

Fig. 2.a depicts the average and 99$^{th}$ percentile of the UL latency (i.e., from the vehicle to the ToD control center) for different cell bandwidths (BW) and 5G configurations (FDD and different TDD)[3]. The figure does not show 99$^{th}$ percentile values for TDD1 and TDD2 with BW<60 MHz since these configurations cannot meet the reliability requirement as the percentage of packets received before the latency limit is below 99%. Packets may not be received because there is a transmission error or because they are dropped at the transmitter. A vehicle would drop a packet if it has not been able to transmit it before a new packet is generated since the ToD service needs the most updated information from the cameras. This situation can occur if, for example, the vehicle does not have access to the necessary radio resources before a new packet is generated. Fig. 2.a shows that the FDD or TDD3 configurations can meet the reliability requirement with a cell bandwidth of only 30 MHz, whereas TDD1 and TDD2 require at least 60 MHz. This is the case because FDD and TDD3 allocate 50% of the radio resources to UL transmissions compared to only 20% in the case of TDD1 and TDD2. The unbalanced UL/DL distribution of radio resources of TDD1 and TDD2 increases their radio latency and hence the latency values reported in Fig. 2.a. For example, the TDD1 configuration increases the average UL radio latency by 140% compared to TDD3 when BW=40 MHz (7.6 ms vs 3.2 ms). On the other hand, both configurations experience nearly the same average latency at the TN, CN, and Internet connection to the ToD control center; the sum of these latencies is approximately equal to 7.6 ms when BW=40 MHz.

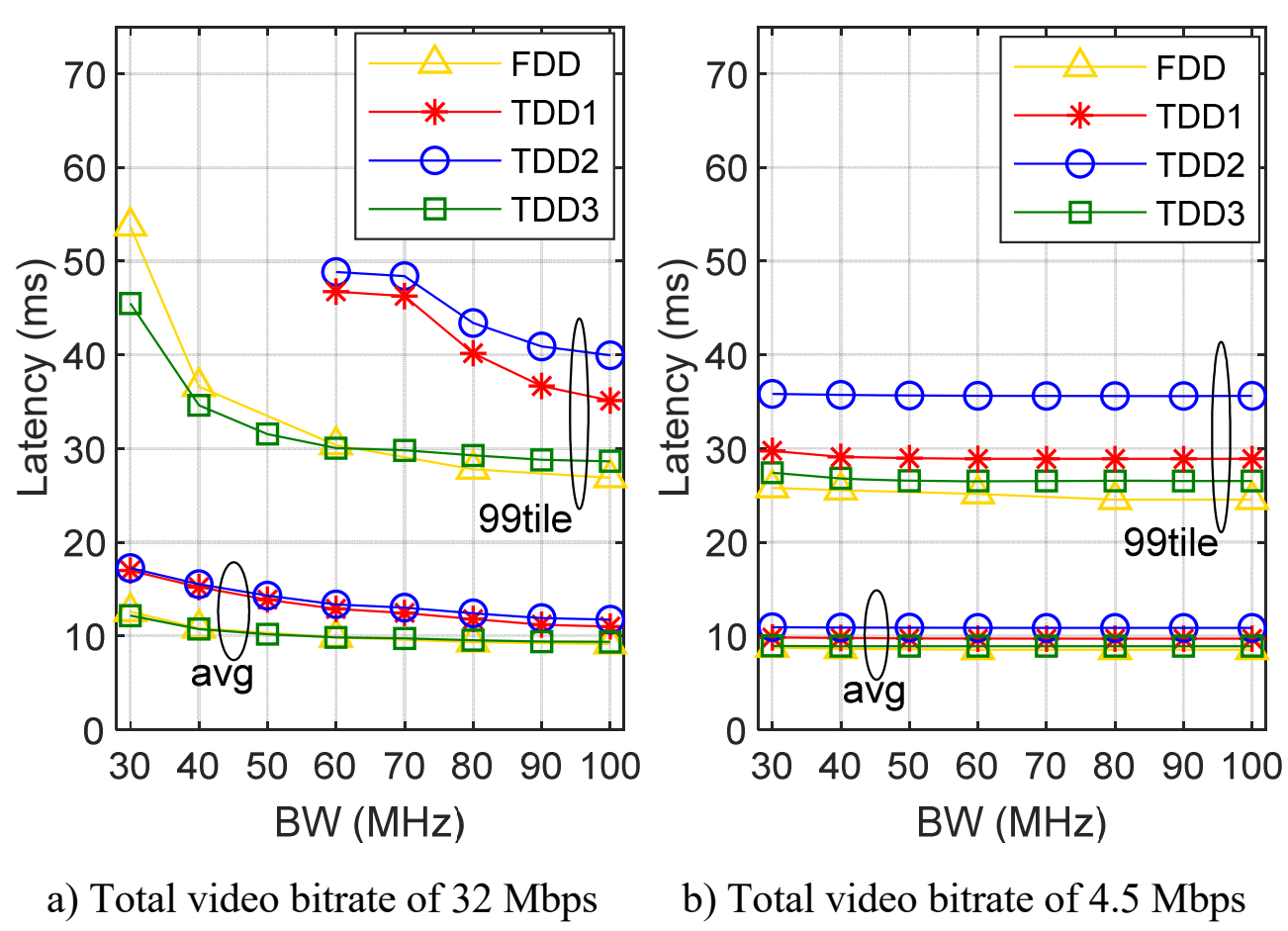

a) Total video bitrate of 32 Mbps b) Total video bitrate of 4.5 Mbps

**Fig. 2**. Average (avg) and 99$^{th}$ percentile of the UL latency.

[3] The average value is calculated considering only the packets received at the ToD control center. The 99$^{th}$ percentile represents the maximum UL latency experienced by the 99% of the packets.

Fig. 2.a shows that the percentage of radio resources allocated to UL transmissions has a stronger impact on the 99$^{th}$ percentile of the UL latency. In particular, the figure shows that the differences between the unbalanced (TDD1 and TDD2) and balanced (TDD3 and FDD) UL/DL configurations increase for the 99$^{th}$ percentile. Fig. 2.a also shows that TDD1 and TDD2 result in different values of the latency for the 99$^{th}$ percentile even though they both allocate 20% of the radio resources to UL transmissions. This is because UL transmissions can only take place every 5 ms with the TDD2 frame structure (DDDDDDDDUU) when considering SCS=30 kHz and a slot time duration of 0.5 ms. On the other hand, UL transmissions can take place every 2.5 ms with the TDD1 frame structure (DDDDU)[4]. Fig. 2.a also shows that the maximum UL latency experienced by 99% of the packets with FDD is slightly larger than using TDD3 when BW<70 MHz despite allocating the same percentage of resources for UL transmissions. This is because the FDD configuration only uses half the bandwidth (BW/2) for UL transmissions. In this case, video frames need to be segmented in a higher number of packets with shorter length than when using TDD3, which increases the radio latency. On the other hand, FDD reduces the latency compared to TDD3 when BW is high because UL transmissions may occur in every slot with FDD compared to in every 2 slots with TDD3 (DUDU).

Fig. 2.a shows that TDD1 and TDD2 can meet the latency and reliability requirements with 60 MHz, whereas TDD3 and FDD only need 30 MHz. However, we should note that the UL latency values reported in Fig. 2.a do not account for processing delays at the ToD control center and human reaction times. The UL latency values in Fig. 2.a are mostly influenced by the radio latency and the latency experienced in the Internet connection between the last UPF of the CN and the ToD control center. For example, the radio latency and the latency introduced by the Internet connection represent 54.4% and 40.2%, respectively, of the 99$^{th}$ percentile UL latency when considering TDD1 and BW=60 MHz. The distance between a vehicle and its serving gNB has also an important impact on the capability to support the ToD service.

Fig. 3 represents the percentage of packets that are not received at the ToD control center as a function of the distance of the vehicle to the gNB for TDD1 and TDD2. Packets may not be received due to transmission errors or because they are dropped in the transmitter. In our analysis, packets can be retransmitted up to three times, and we adapt the MCS per packet based on the experienced CQI to guarantee a BLER of 10% [13]. In this case, the percentage of packets received with error remains consistently at 0.01% for all distances between the gNB and the vehicle. The variations observed in Fig. 3 are then due to an increase of the packets dropped in the transmitter with the distance. Propagation conditions degrade with the distance to the gNB, and vehicles need to use more robust MCSs to strengthen the error protection. In this case, vehicles require a higher number of radio resources to transmit a packet, and vehicles might need to wait longer to have access to the necessary resources to transmit the packet. This increases the risk of packets being dropped and explains why the percentage of packets that are not received increases with the distance in Fig. 3. Following the trends depicted in Fig. 3, the ToD reliability requirement cannot be met for distances greater than 551 m when BW=30 MHz. If BW increases to 50 MHz, it is possible to guarantee the reliability requirement up to a distance of 787 m.

The analysis presented so far considers the ToD service defined by the 5GAA with a total video bitrate of 32 Mbps from 4 cameras [2]. Trials conducted in European projects [10][16] have shown that ToD can be supported using lower resolution video. For example, [10] deploys a ToD service with a total video bitrate of 4.5 Mbps generated by 4 cameras. Fig. 4 analyses the impact of the total video bitrate on the percentage of received UL packets. Results are depicted for TDD1 as a function of BW, but similar trends are observed with TDD2. We focus this analysis on TDD1 and TDD2 since they are used in most current 5G deployments, and they only allocate 20% of radio resources to the UL. Fig. 4 shows that the bandwidth needed to support the UL ToD reliability requirement (99% of packets received) decreases from 60 to 30 MHz with video bitrates equal to or lower than 14 Mbps. Fig. 2.b depicts then the latency performance achieved with a total video bitrate of 4.5 Mbps for all different FDD and TDD configurations. The comparison of Fig. 2.a and Fig. 2.b shows that reducing the video bitrate considerably decreases the UL latency, especially for lower cell bandwidths. It is now possible to satisfy the reliability requirement with only 30 MHz when using TDD1 and TDD2.

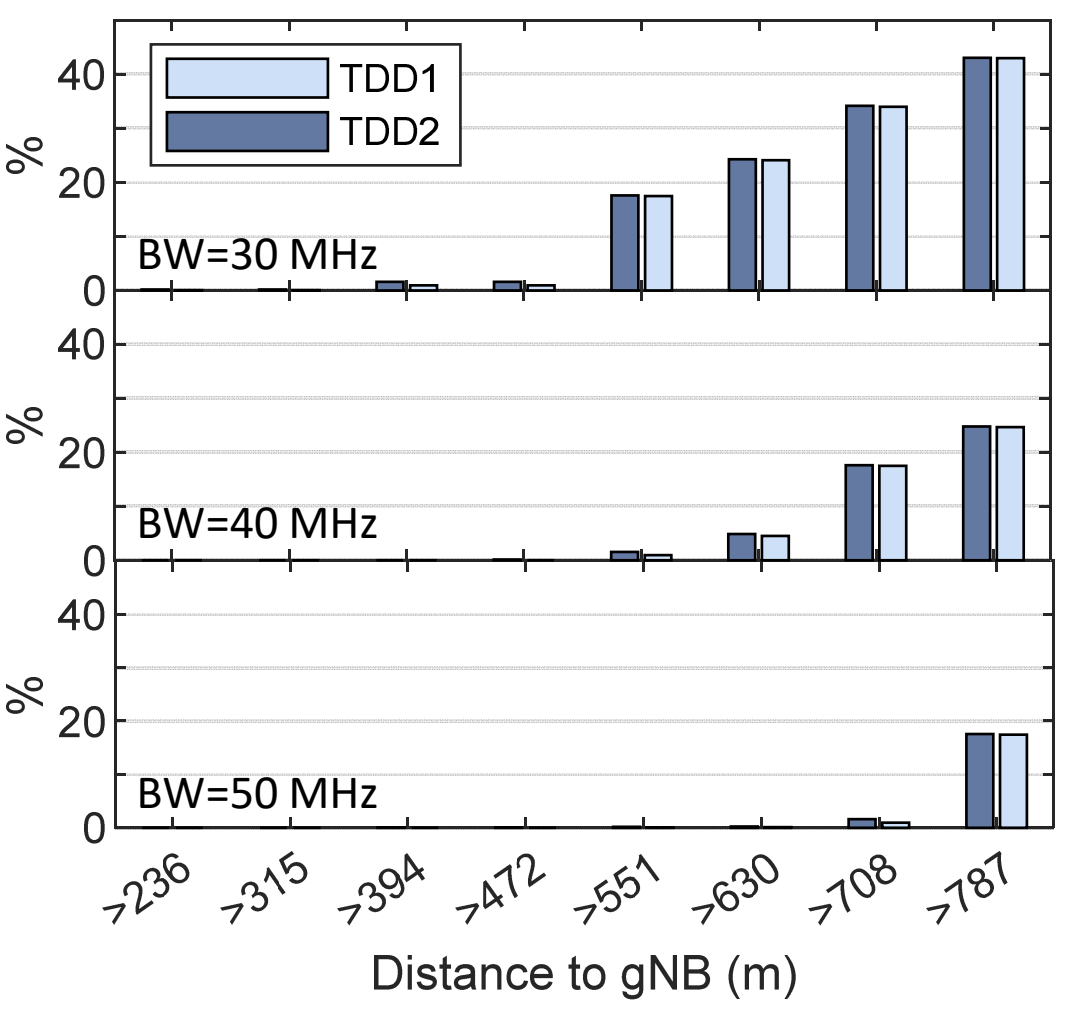

**Fig. 3**. Percentage of packets that are not received at the ToD control center as a function of the distance of the vehicle to the gNB.

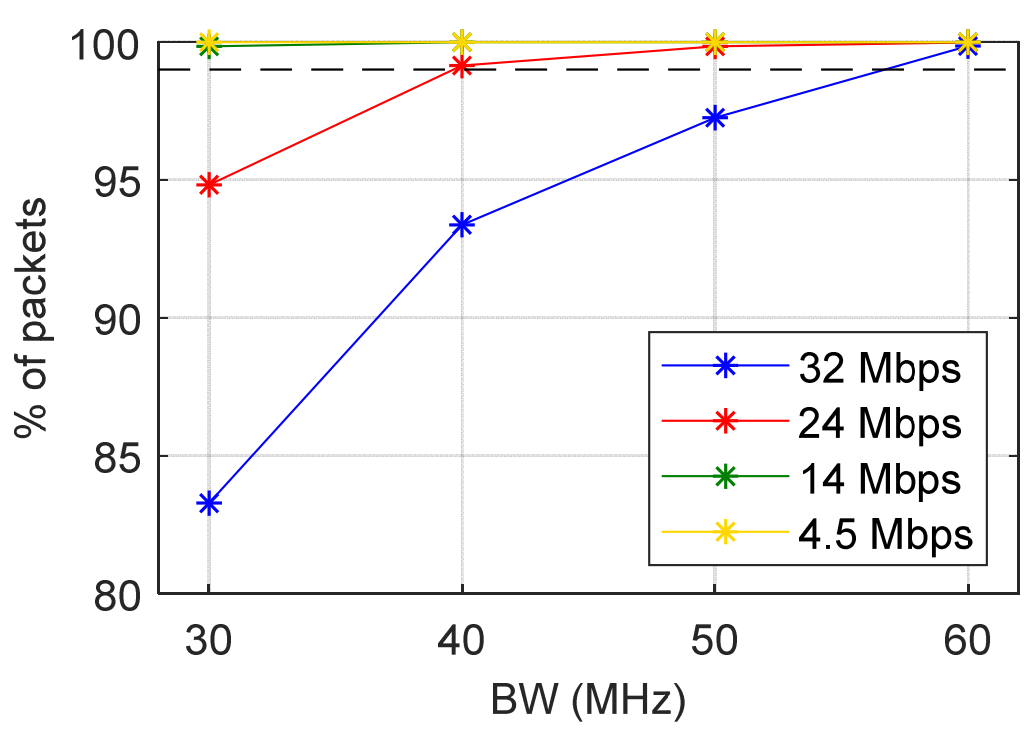

**Fig. 4**. Percentage of UL packets received as a function of the cell bandwidth and the video bitrates for TDD1.

[4] The differences between TDD1 and TDD2 get progressively larger when the video frame generation rate increases, i.e., when the video inter-frame space decreases. We evaluated the performance achieved for increasing video frame generation rates and observed that the percentage of received packets decreases more for TDD2 than TDD1 because of the larger time between UL slots.

The results in Fig. 2.b show that the latency achieved with the lowest video bitrate is not affected by the cell bandwidth when there is a single ToD vehicle per cell. The latency experienced in the TN, CN and the Internet connection is almost constant for all BW values (7.54 and 21.1 ms for the average and 99$^{th}$ percentile latency, respectively). The radio latency is also not affected by the cell bandwidth because video frames at 4.5 Mbps are transmitted in a single packet over the radio channel (i.e. video frames are not segmented) even when BW=30 MHz. TDD1 and TDD2 result in higher average and 99$^{th}$ percentile values of the UL latency than FDD or TDD3 because vehicles need to wait more time between UL slots. For example, a vehicle waits on average 1 slot (of 0.5 ms with SCS 30 kHz) for the first radio transmission of a packet when using TDD3. If the packet needs to be retransmitted (with a 0.1 probability in our scenario), the latency increases by 4 ms for each retransmission with TDD3 because of the signalling exchange needed to request a retransmission and gain access to the radio resources. If a vehicle utilizes TDD1 or TDD2, it needs to wait on average 2.5 and 5 slots, respectively, to transmit a packet, and the average radio latency increases by 5 and 10 ms for each retransmission if TDD1 and TDD2 are used respectively. These trends explain the differences between TDD and FDD configurations observed in Fig. 2.b.

### B. *DL ToD traffic*

A ToD vehicle receives commands from the remote ToD control center to control its movement. 5GAA establishes that these DL commands (1000 bytes) should be sent every 20 ms and need to be received in less than 20 ms with 99.999% reliability [2]. These commands are independent of the video bitrates sent from the vehicle. Fig. 5 depicts the maximum DL latency experienced by 99.999% of the packets for all FDD and TDD configurations when BW=40 MHz. Fig. 5 shows the contribution of each latency component to the DL latency. The figure shows that the latency requirement established by the 5GAA (i.e., 20 ms for DL, see Table I) is too demanding and cannot be met whatever the 5G configuration when considering centralized 5G network deployments where the ToD control center is connected through the Internet. This is independent of the BW allocated since Fig. 5 shows that the radio latency (between 1.5 and 2 ms with all FDD and TDD configurations) contribution is minimal even with BW=40 MHz. The DL latency is actually dominated by the latency at the Internet connection between the CN and the V2X-AS (over 50 ms).

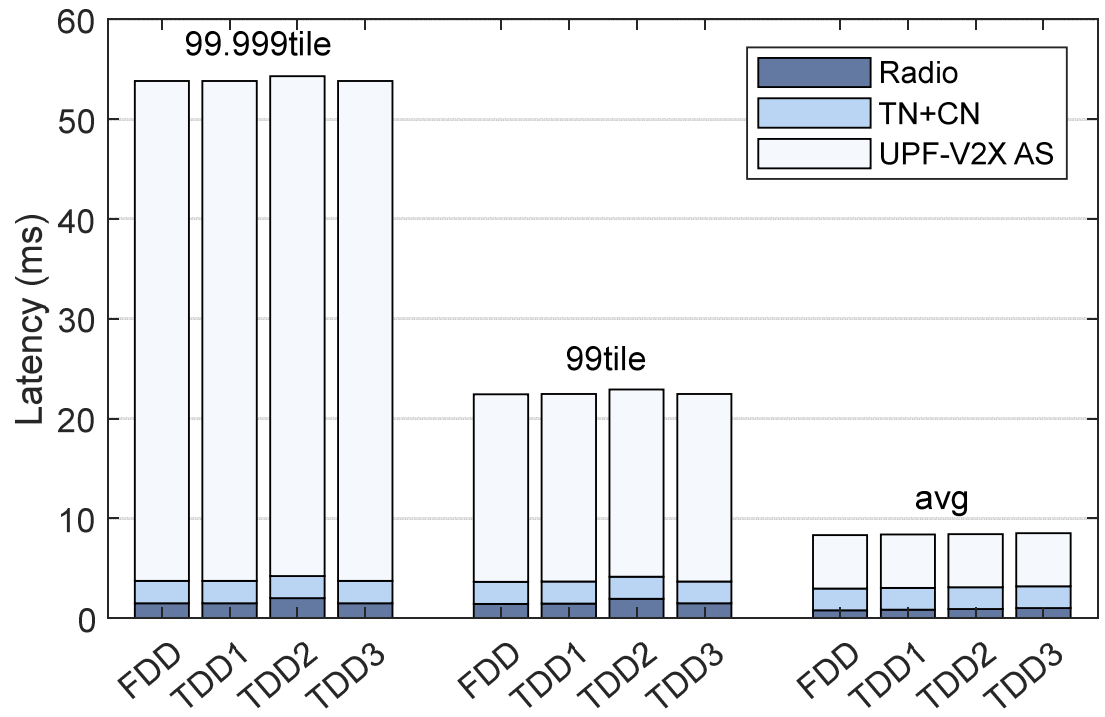

**Fig. 5**. 99.999th (99.999tile) and 99th (99tile) percentiles and average (avg) of the DL latency. The figure shows the latency contributions of the radio network, TN and CN, and the Internet connection between the last UPF of the CN and the V2X-AS. BW=40 MHz.

We extend the analysis in Fig. 5 to consider a more relaxed 99% reliability requirement. The 99$^{th}$ percentile of the DL latency is still dominated by the Internet latency even if it significantly reduces as we relax the reliability requirement. The figure shows that all TDD and FDD configurations achieve similar performance (close to the 20ms latency requirement). This is because all configurations have sufficient DL resources to easily accommodate the traffic load generated by the commands; in fact, no packets are dropped with any of the configurations.

## VI. SCALABILITY

Previous results have shown that 5G networks with centralized deployments might be able to support the ToD UL traffic depending on the bandwidth available, the video bit rates, and the utilized TDD frame structures. The capacity to support the DL commands is constrained by the stringent ToD requirements and the latency of the Internet connection to the ToD control center. The previous section focused on a single ToD vehicle per cell, and this section extends the study to several ToD vehicles per cell.

The possibility to support several ToD vehicles per cell is limited if each vehicle transmits video at 32 Mbps. Out of all configurations analyzed, it is only possible to support the 99% reliability requirement for UL ToD traffic for a density of 1 veh/km/lane, a bandwidth of 100 MHz and the TDD3 and FDD configurations. In this case, the maximum UL latency experienced by 99% of the packets (or 99$^{th}$ percentile of the UL latency) is below the 100 ms requirement (40.3 ms for TDD3 and 50.4 ms for FDD). All other configurations (duplexing and BW) cannot guarantee receiving 99% of the UL packets in less than 100 ms. If we increase the density to 2 veh/km/lane, the FDD and TDD3 configurations cannot guarantee it even with BW=100 MHz. In this case, the percentage of packets that are received in less than 100 ms drops to 55% for TDD3 and 52% for FDD.

Scaling the ToD service to the levels identified by 5GAA in [2] (i.e., 10 veh/km$^2$, see Table I) requires vehicles to reduce their total video bitrate. Fig. 6 depicts the 99th percentile of the UL latency as a function of BW when each vehicle transmits a total bitrate of 4.5 Mbps. Results are represented for two densities of ToD vehicles. The figure shows that most configurations (except TDD1 and TDD2 with BW=30 MHz) can guarantee the 99% reliability requirement and an UL latency below 100 ms when the density of ToD vehicles is equal to 1 veh/km/lane. Increasing the density of ToD vehicles to 2 veh/km/lane decreases the possible duplexing and BW configurations that meet the reliability and latency requirements (Fig. 6.b). Only the TDD3 configuration can sustain this density of ToD vehicles when BW is lower than 60 MHz. TDD1 and TDD2 require at least 80 MHz for satisfying the 99% reliability requirement and achieving a 99$^{th}$ percentile UL latency below 100 ms.

5GAA currently requires that 99.999% of the commands from the ToD control center to the vehicle are received in less than 20 ms. Fig. 7 depicts the maximum DL latency experienced by 99.999% of the DL packets when there are 1 and 2 veh/km/lane and BW=40 MHz. Under these configurations, no packets are dropped at the transmitter because the network has sufficient radio resources to support the DL traffic load even when the density of ToD vehicles increases to 2 veh/km/cell. Fig. 7 shows that the sum of latency experienced in the radio network, TN and CN is lower than 7.3 ms with FDD and all TDD configurations even for a density of ToD vehicles equal to 2 veh/km/lane. However, the 20 ms latency requirement cannot be met due to the latency introduced by the Internet connection to the ToD control center (equal to 50.1 ms for the 99.999 percentile).

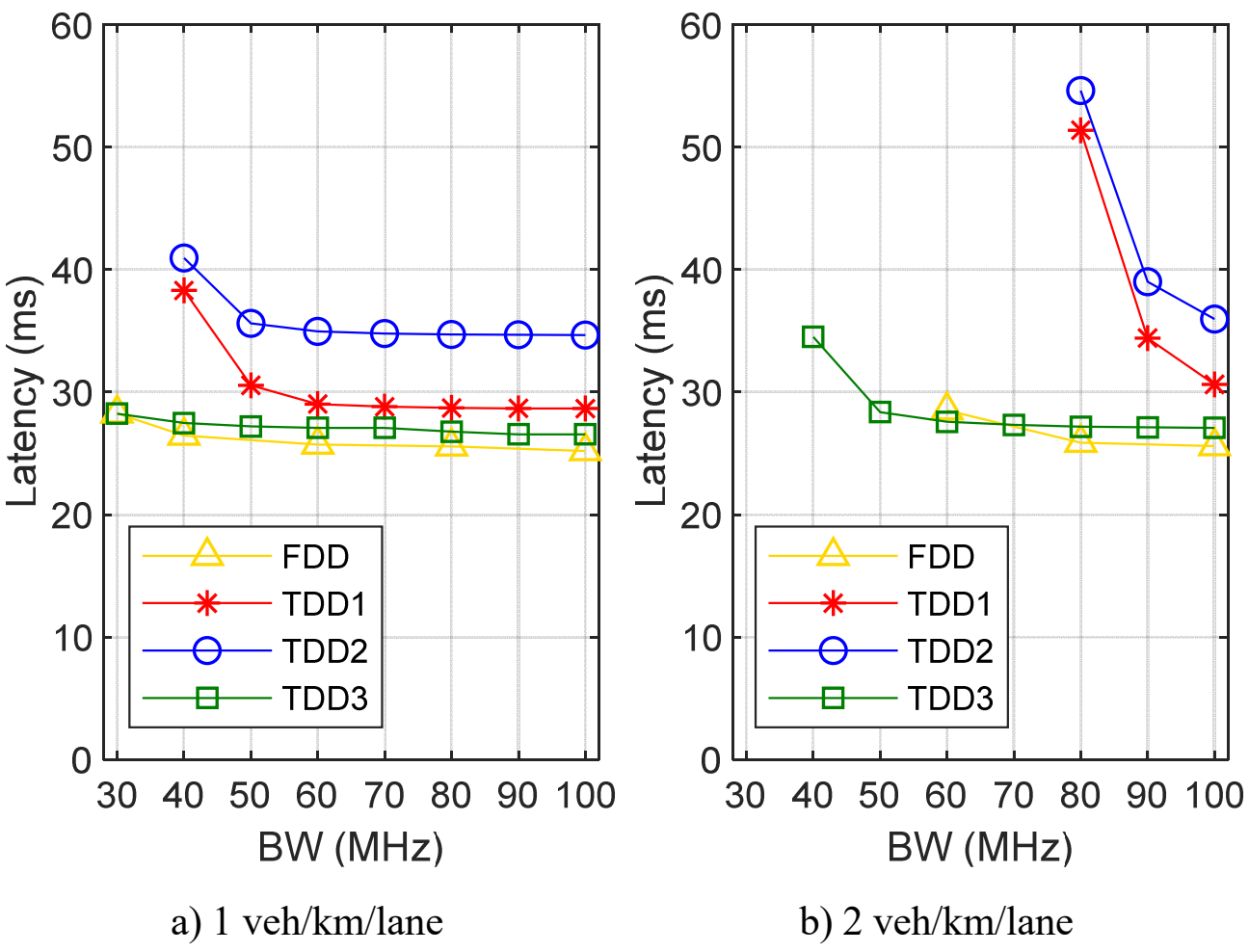

**Fig. 6**. 99th percentile of the UL latency as a function of BW. Vehicles transmit video at a total bitrate of 4.5 Mbps.

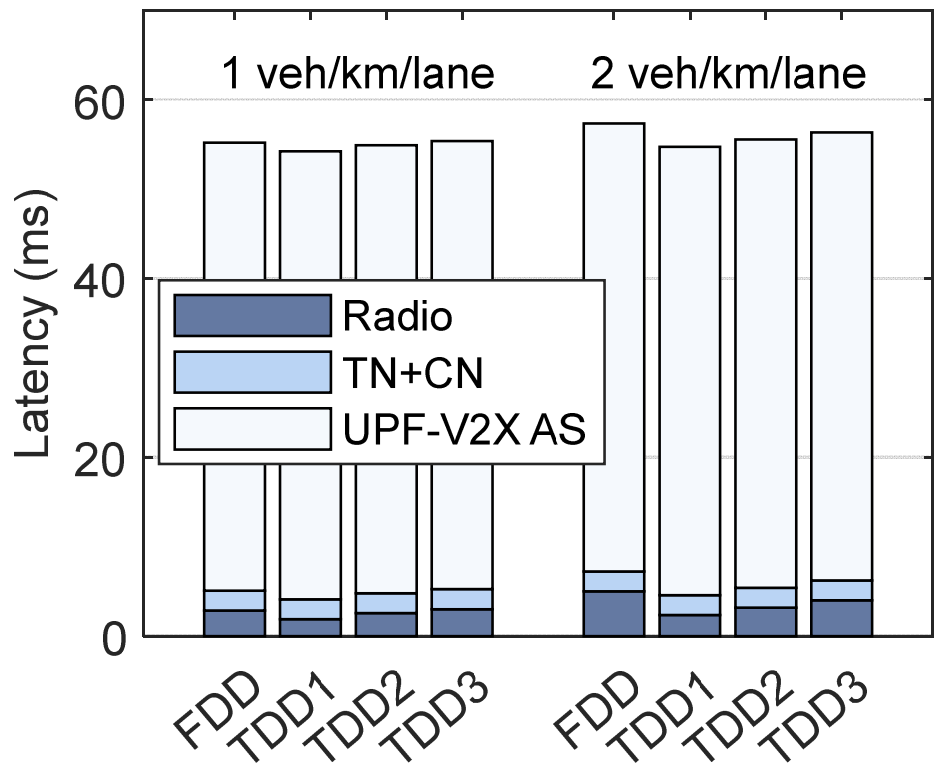

**Fig. 7**. 99.999th percentiles of the DL latency for BW=40 MHz and vehicle densities of 1 and 2 ceh/km/lane.

## VII. Conclusions

This study has analyzed the capability of 5G networks to support teleoperated driving under a centralized network deployment, and different 5G configurations and densities of ToD vehicles. The study has demonstrated that commonly utilized 5G TDD frame structures require more than 60 MHz to support the ToD service specified by 5GAA even when there is a single ToD vehicle per cell. This is due to the unbalanced distribution of slots between UL and DL. The bandwidth required can be reduced to 30 MHz using more balanced TDD frame structures, an FDD mode, or reducing the video bitrates uploaded from the ToD vehicles. Our scalability study implies that reducing the video bitrate is necessary for supporting the target density of ToD vehicles identified by the 5GAA. Supporting the very stringent requirements established by the 5GAA to transmit ToD commands from the remote center to the vehicles requires transitioning from centralized to Multi-Access Edge Computing (MEC)-based network deployments due to the latency introduced by the Internet connection to the ToD control center.

## Acknowledgment

This work was supported in part by MCIN/AEI/10.13039 /501100011033 (grants IJC2018-036862-I, PID2020-115576RB-I00), by the "European Union NextGenerationEU/PRTR" (TED2021-130436B-I00), and by Generalitat Valenciana (CIGE/2022/17), and UMH's Vicerrectorado de Investigación grants.